\RequirePackage{amssymb,amsmath}
\documentclass[a4paper,runningheads]{llncs}
\usepackage{verbatim}
\usepackage{epsfig}

\usepackage[utf8]{inputenc}
\usepackage[T1]{fontenc} % TEMP
\DeclareSymbolFont{rsfscript}{OMS}{rsfs}{m}{n}
\DeclareSymbolFontAlphabet{\mathrsfs}{rsfscript}

\spnewtheorem{fact}{Fact}{\bfseries}{\itshape}

\newcommand{\keywords}[1]{\par\addvspace\baselineskip
\noindent\keywordname\enspace\ignorespaces#1}

\usepackage{geometry,array,graphicx}
\usepackage{soul}
\usepackage{color}
\usepackage[usenames,dvipsnames]{xcolor}
\usepackage{algorithm}
\usepackage{algpseudocode}

\usepackage{graphicx}

\newcommand{\E}[0]{\mathbb{E}} 
\newcommand{\Ex}{\mathbb{E}}
\renewcommand{\Pr}{\mathbb{P}}
\newcommand{\<}{\langle}
\renewcommand{\>}{\rangle}

\begin{document}
\title{A Fast Algorithm Finding the Shortest Reset Words}
%\author{Andrzej Kisielewicz\inst{1} \and Jakub Kowalski\inst{2} \and Marek Szyku{\l}a\inst{2}}

%\institute{Institute of Mathematics, University of Wrocław, pl. Grunwaldzki 2/4, 50-384 Wrocław, Poland\\ e-mail: \email{kisiel@math.uni.wroc.pl}
%\and Institute of Computer Science, University of Wrocław, ul. Joliot-Curie 15, 50-383 Wrocław, Poland\\ e-mail: \email{\{kot,msz\}@ii.uni.wroc.pl}}

\author{Andrzej Kisielewicz\thanks{Supported in part by Polish MNiSZW grant N N201 543038.}\inst{1,2} \and Jakub Kowalski\inst{1} \and Marek Szyku{\l}a\inst{1}}

\institute{Department of Mathematics and Computer Science, University of Wrocław, Poland \and Institute of Mathematics and Computer Science, University of Opole, Poland \\  \email{andrzej.kisielewicz@math.uni.wroc.pl,\  \{kot,msz\}@ii.uni.wroc.pl}}

\maketitle

\begin{abstract}

In this paper we present a new fast algorithm for finding minimal reset words for finite synchronizing automata, which is a problem appearing in many practical applications. The problem is known to be computationally hard, so our algorithm is exponential in the worst case, but it is faster than the algorithms used so far and it performs well on average. The main idea is to use a bidirectional BFS and radix (Patricia) tries to store and compare subsets. Also a number of heuristics are applied. We give both theoretical and practical arguments showing that the effective branching factor is considerably reduced. As a practical test we perform an experimental study of the length of the shortest reset word for random automata with $n\leq 300$ states and 2 input letters. In particular, we obtain a new estimation of the expected length of the shortest reset word $\approx 2.5\sqrt{n-5}$.

\keywords{Synchronizing automaton, synchronizing word, \v{C}ern\'{y} conjecture}
\end{abstract}

%%%%%%%%%%%%%%%%%%%%%%%%%%%%%%%%%%%%%%%%%%%%%%%%%%%%%%%%%%%%%%%%%%%%%%%%%%%%%%%%%%%%%%

\section{Introduction}
We deal with (complete deterministic) finite automata  $A = \< Q,\Sigma,\delta \>$ with the state set $Q$, the input alphabet
$\Sigma$, and the transition function $\delta : \; Q \times \Sigma \to Q$. The action of $\Sigma$ on $Q$ given by $\delta$ is denoted simply by
concatenation:  $\delta(q,a) = qa$. This action extends naturally to the
action  $qw$ of words for any $w\in \Sigma^*$. If $|Qw|=1$, that is, the image of $Q$ by $w$ 
consists of a single state, then $w$ is called a \emph{reset} (or \emph{synchronizing}) word for $A$, and $A$ itself is called \emph{synchronizing}. (In other words, $w$ resets (synchronizes) $A$ in the sense that, under the action of $w$, all the states are sent into the same state). The synchronizing property is very important, because it makes the automaton resistant to errors that could occur in an input word. After detecting an error a synchronizing word can be used to reset the automaton to its initial state. Synchronizing automata have many practical applications. They are used in robotics (for designing so-called part orienters) \cite{AV2003}, bioinformatics (the reset problem) \cite{BAPLS2003}, network theory \cite{Ka2002}, theory of codes \cite{Ju2008} etc. 

Theoretical research in the area is mainly motivated by the \v{C}ern\'{y} conjecture stating that every synchronizing automaton $A$ with $n$ states has a reset word of length $\leq (n-1)^2$. This conjecture was formulated by \v{C}ern\'{y} in 1964 \cite{Ce1964}, and is considered the most longstanding open problem in the combinatorial theory of finite automata. So far, the conjecture has been proved only for a few special classes of automata and a general cubic upper bound $(n^3-n)/6$ has been established (see Volkov \cite{Vo2008} for an excellent survey of the results, and Trahtman \cite{Tr2011} for a recently found new cubic bound). Using computers the conjecture has been verified for small automata with 2 letters and $n\leq 10$ states (and with $k\leq 4$ letters and  $n\leq 7$ states \cite{Tr2006}; see also \cite{AGV2010} for $n=9$ states). It is known that, in general, the problem is computationally hard, since it involves an NP-hard decision problem. Recently, it has been shown that the problem of finding the length of the shortest reset word is $\mathrm{FP^{NP[log]}}$-complete, and the related decision problem is both NP- and coNP-hard \cite{OM2010}. 

On the other hand, there are several theoretical and experimental results showing that most synchronizing automata have relatively short reset words and those slowly synchronizing (with the shortest reset words of quadratic length) are rather exceptional \cite{AGV2010}. An old result by Higgins \cite{Hi1988} on products in transformation semigroups shows that a random automaton with an alphabet of size larger than $2n$ has, with high probability, a reset word of length $\leq 2n$. More recently, it was proved that, for every $\epsilon > 0$, a random automaton with $n$ states over an alphabet of size $n^{0.5+\epsilon}$, with high probability, is synchronizing and satisfies the \v{C}ern\'{y} conjecture \cite{SZ2010}. In computing reset words, either exponential algorithms finding the shortest reset words \cite{ST2011,Tr2006,KRW2012} or polynomial heuristics finding relatively short reset words \cite{GH2011,KRW2012,Ro2005,Ro2009,Tr2006} are widely used. 
The standard approach is to construct the power automaton and to compute the shortest path from the whole set state to a singleton \cite{Sa2005,Tr2006,KRW2012,Vo2008}. Most naturally, the breadth-first-search method is used which starts from the set of all states of the given automaton and forms images applying letter transformations until a singleton is reached. Based on these ideas computation packages have been created (TESTAS \cite{Tr2003} and recently developed COMPAS \cite{CR2011}). In \cite{Ro2009}, Roman uses a genetic algorithm to find a reset word of randomly generated automata and thus obtains upper bounds on the length of the shortest reset word.

A new interesting approach for finding the exact length using a SAT-solver has been applied recently by Skvortsov and Tipikin \cite{ST2011}. The problem of determining if an automaton has a reset word of length at most $l$ is reduced to the SAT problem and the binary search for the exact length is performed. Using this approach, the following experimental study is done. For chosen numbers $n$ of states from the interval $[1,100]$ random automata with 2 input letters are generated, checked if they are synchronizing, and if so, the shortest reset word is computed. The results directly contradict the conjecture made by Roman \cite{Ro2009} that the mean length of the shortest reset word for a random $n$-state synchronizing automaton is linear and almost equal to $0.486n$. Skvortsov and Tipikin argue that their experiment based on a larger set of data shows that this length is actually sublinear and $\approx 1.95n^{0.55}$. 

In this paper we present a new algorithm based on a bidirectional breadth-first-search. Implementing this idea requires efficiently solving the problem of storing and comparing resulted subsets of states. To this aim radix tries (also known as Patricia tries \cite{Mo1968}) are used. We analyze the algorithm from both theoretical and practical sides. As the first test of efficiency we have performed experiments analogous to those done by Skvortsov and Tipikin. Due to the well performance of the algorithm we were able to generate and check one million automata for each $n \leq 100$, (compared with $200$--$2000$ generated by Skvortsov and Tipikin), and we were able to test much larger automata with up to $n= 320$ states.  
Our data confirm the hypothesis that the expected length of the shortest reset word is sublinear, but show that more precise is a smaller approximation  $\approx 2.5\sqrt{n-5}$. In addition, the larger set of data enables us to estimate the error and to show that for our approximation with high probability the error is very small. We also verify and discuss other results and claims of \cite{ST2011}. 

Our algorithm makes also possible to find a reset word of the shortest length (not only the length). Curiously, it works in polynomial time for known slowly synchronizing automata series \cite{AGV2010}. So far, most of the empirical research in the area concerns automata with 2 input letters. Some  researches suggest that automata with more letters may exhibit a different behavior. We plan to use the algorithm to perform an extensive research on automata with $k>2$ letters.

\section{Algorithm}

The algorithm gets an automaton $A = \< Q,\Sigma,\delta \>$ with $n$ states and $k$ input letters. First, $A$ is checked if it is synchronizing using the well known (and efficient) algorithm \cite{Ep1990}. If so, then we proceed to search for a synchronizing word of the shortest length. Here, one may perform the breadth-first search (BFS) on the power automaton of $A$ starting from the set $Q$ of all the states and computing successive images by the letters of the alphabet $\Sigma$ (and recording the sequences of the letters applied). One may also search in the inverse (backward) direction starting from the singleton sets and computing successive preimages (this search will be refereed to as IBFS). Both the searches have branching factor $k$ (the number of input letters) and need to compute $O(k^l)$ sets (or $O(nk^l)$ in IBFS) to find a synchronizing word of the shortest length $l$. The idea behind bidirectional search is to perform two searches simultaneously and check if they meet. Then a synchronizing word may be found in only $O(nk^{l/2})$ steps. However, to implement this idea there must be an efficient way to check each new subset to see if it already appears in the search tree of the other half of the search.

\subsection{General Ideas}

For each search we maintain the current list of subsets that can be obtained from the start in a given number of steps. Since the lists have a tendency to grow exponentially and to contain subsets obtained on earlier steps, it is more efficient to maintain additional lists of visited subsets (for each search) and to use them to remove from the current lists redundant subsets. We have checked experimentally that it is a good strategy to decrease the branching factor.

To check if the two searches meet one needs to perform \emph{subset checking}: after each step, BFS or IBFS, we check if a set on the current IBFS list \emph{contains} a set on the current BFS list. If so, it means that there are words $u,w \in \Sigma^*$ such that the image $Qu$ is a subset of the preimage $\{q\}w^{-1}$ for some $q\in Q$. Consequently, $Quw = \{q\}$, as required.

Since, in the bidirectional approach, subset checking must be performed anyway, it may be also applied to reduce lists using the following simple observation. If $S$ and $T$ are subsets of $Q$ such that $S \subseteq T$, then $|Tw|=1$ implies $|Sw|=1$ for any $w\in\Sigma^*$.
It follows that, for example, a subset on the IBFS list contains a subset on the BFS list if and only if -- with respect to inclusion -- a maximal element on the IBFS list contains a minimal element on the BFS list. Consequently, the only subsets on the BFS lists we need to consider are those minimal with respect to inclusion and the only subsets on the IBFS lists we need to consider are those maximal with respect to inclusion.

To store and check subsets on the lists we apply an efficient data structure known as \emph{radix trie} (Patricia trie) \cite{Mo1968}. We show that the \emph{subset checking} operation (checking whether a given set $S$ has a subset stored in the trie) and the dual \emph{superset checking} (checking whether a given set $S$ has a superset stored in the trie) are efficient enough for these structures to make a combination of the ideas presented above work well in practice.

This approach is fast but memory consuming.
In order to also make the algorithm work efficiently for larger automata, when the memory limit is reached, the bidirectional approach is replaced by a sort of an inverse DFS search not involving the tries of visited subsets anymore. We also apply several technical optimizations and heuristics which yields a considerable speed-up. They are described in Section~\ref{sec_heuristics}.

\subsection{Radix Tries}
A \emph{radix trie} is a binary tree of the maximal depth $n$ which stores subsets of a given $n$-set $Q$ in its leaves. Having a fixed linear order of elements $q_1,\ldots,q_n \in Q$, each subset $S$ of $Q$ encodes a path from the root to a leaf in the natural way: after $i$ steps the path goes to the right child whenever $q_i \in S$, and goes to the left, otherwise. A radix trie is \emph{compressed} in the sense that instead in a node at depth $n$ it stores a subset in the first node that determines uniquely the subset in the stored collection (no other subset shares the same path as a prefix of the encoding); c.f. \cite{Mo1968}.

The insert operation for radix tries is natural and can be performed in at most $n$ steps. The \emph{subset checking} operation is performed by a depth-first-search checking if the given set $S \subseteq Q$ contains a subset stored in the visited leaf. An essential advantage is that the search does not need to branch into the right child of a node if the checked subset $S$ does not contain the state corresponding to the current level. The superset checking operation (for IBFS) is done in the dual way. These issues are discussed in more detail in \ref{subsec_bidirectdesc}.

\begin{algorithm}\caption{The main part}\label{alg_bidirect}
\algrenewcommand\algorithmicrequire{\textbf{Input}}
\begin{algorithmic}[1]
\Require $A = \langle Q,\Sigma,\delta \rangle$ -- a synchronizing automaton with $n=|Q|$ states and $k =|\Sigma|$ input letters.
\Require $\mathtt{maxlen}$ -- maximum length of words to be checked. 
\Statex \Comment{Initialize four radix tries to store and handle subsets of $Q$:} 
\State $T_{c}  \gets$ \Call{EmptyTrie}{} \Comment{BFS current trie}
\State $T_{v}  \gets$ \Call{EmptyTrie}{} \Comment{BFS visited trie\; }
\State $T_{ic} \gets$ \Call{EmptyTrie}{} \Comment{IBFS current trie}
\State $T_{iv}  \gets$ \Call{EmptyTrie}{} \Comment{IBFS visited trie\; }
\State \Call{$T_{c}.$insert}{$Q$}
\State \Call{$T_{v}.$insert}{$Q$}
\ForAll{$q \in Q$}
  \State \Call{$T_{ic}.$insert}{$\{q\}$}
  \State \Call{$T_{iv}.$insert}{$\{q\}$}
\EndFor
\For{$l \gets 1$ \textbf{to} $\mathtt{maxlen}$} 
  \If{estimated time of the BFS step is smaller than that of IBFS} 
    \State \Call{BFS\_Step}{$T_{c}$,$T_{v}$}   \Comment{Modify BFS tries; minimize $T_c$ using $T_v$} 
  \Else
    \State \Call{IBFS\_Step}{$T_{ic}$,$T_{iv}$}   \Comment{Modify IBFS tries; minimize $T_{ic}$ using $T_{iv}$}  
  \EndIf
  \ForAll{$S \in T_{ic}$}\Comment{The goal test loop}
    \If{\Call{$T_{c}.$contains\_subset\_of}{$S$}}
      \State \Return $l$ \Comment{The length of the shortest reset word}
    \EndIf
  \EndFor
\EndFor
\State \Return ''No synchronizing word of length $\leq \mathtt{maxlen}$''
\end{algorithmic}
\end{algorithm}

\subsection{Description}\label{subsec_bidirectdesc}

The main part of the algorithm is given in Algorithm~\ref{alg_bidirect}. To make it clearer we restrict the task to finding the \emph{shortest length} of a reset word only. Yet, the algorithm can be easily modified to return also a reset word of such length (see \ref{subsec_words}). 

We use, in principle, four radix tries $T_{c}, T_{v}, T_{ic}, T_{iv}$ to maintain the BFS current, BFS visited, IBFS current, and IBFS visited lists, respectively. After initializing the tries we enter a loop consisting of at most $\mathtt{maxlen}$ steps (line~11). In each step we perform a step of the BFS procedure or IBFS procedure depending on comparison of estimated expected execution time of both steps, which we discuss in \ref{subsec_eststep}.

With no regard if BFS or IBFS step was performed recently, in lines~17-21 of Algorithm~\ref{alg_bidirect}, the same goal test loop is performed. For each $S$ in $T_{ic}$, the procedure $T_c.\Call{contains\_subset\_of}{S}$ is executed, which checks if $T_c$ contains a subset of $S$. If so, we claim that $l$ is the shortest length of a rest word for $A$. To prove this we need to analyze the content of the BFS and IBFS steps.

In BFS step (Algorithm~\ref{alg_bfsstep}), for each set $S'$ in the current BFS trie and for each input letter $a$ we compute the image $S=S'a$ and insert it to the list $L$. For each set $S\in L$ we check if a subset of $S$ is already in the BFS visited trie. If so, we skip it. If not, we insert $S$ into the BFS visited trie and in the (newly formed; line~9) BFS current trie $T_c$. Processing elements of $L$ (line~10) in \emph{ascending cardinality order} is a heuristic aimed in getting more subsets skipped in the checking subset procedure in line~11, and in consequence, to deal with smaller structures. It also guarantees that $T_c$ contains only minimal sets in terms of inclusion (the proof of this fact and all other proofs will be given in the extended version of this paper).

After executing lines~10-15 of Algorithm~\ref{alg_bfsstep} the trie $T_v$ may contains some redundant subsets (which are not minimal with respect to inclusion). Therefore in lines~16-18 we have an additional procedure to reduce $T_v$ completely.

The procedure \Call{$T_{v}$.reduce}{} consists of two steps. First, we form a list of elements of $T_v$ using a DFS-search from the left to the right (smaller subsets first). This guarantees that if $S$ precedes $T$ on the list then $S$ does not contain $T$. Hence the only pairs of comparable elements on the list are those with $S$ preceding $T$ and $S \subset T$. In the second step we insert the elements from the list into the empty $T_v$ depending on the result of subset checking performed before each insertion. This guarantees that if a subset $S$ of $T$ is inserted then $T$ will be skipped on the later step. Hence the resulting trie $T_v$ contains no comparable subsets, as required.

Unfortunately, this procedure applied for such a large trie as $T_v$ (which may be of exponential size in terms of $n$) may be time-consuming. We found experimentally that if the trie has not grown too large since the last reduction it is more effective to process a larger trie rather than to perform reduction. In our implementation we perform it after the first step and then only when $T_v$ contains at least $k$ times more sets since it had after the last reduction (which is the worse case for one step with branching factor $k=|\Sigma|$).  

The IBFS step is dual and completely analogous. In line~10 ascending cardinality order is replaced by descending one, in line~5 we compute preimages instead of images, and in line~11 subset checking is replaced by superset checking.

\begin{algorithm}\caption{BFS step procedure}\label{alg_bfsstep}
\begin{algorithmic}[1]
%% \Require $A = \langle Q,\Sigma,\delta \rangle$ -- an automaton.
%% \Require $T_{c},T_{v}$ -- the current and the visited trie, which will be modified.
\Procedure{BFS\_Step}{$T_{c}$,$T_{v}$}
\State $L \gets$ \Call{EmptyList}{} \Comment{The list of all new images}
\ForAll{$S' \in T_{c}$}
  \ForAll{$a \in \Sigma$}
    \State $S \gets \delta(S',a)$ \Comment{Compute the image of $S'$ by the letter $a$}
    \State \Call{$L$.insert}{$S$} 
  \EndFor
\EndFor
\State $T_{c} \gets$ \Call{EmptyTrie}{}
\ForAll{$S \in L$ in ascending cardinality order}
  \If{\textbf{not} \Call{$T_{v}$.contains\_subset\_of}{$S$}}
    \State \Call{$T_{v}$.insert}{$S$}
    \State \Call{$T_{c}$.insert}{$S$}
  \EndIf
\EndFor
\If{$T_{v}$ has grown large since the last reduction}
  \State \Call{$T_{v}$.reduce}{}
\EndIf      
\EndProcedure
\end{algorithmic}
\end{algorithm}

One can prove the following

\begin{theorem}\label{th_bidirectcorrect}
Given a synchronizing $n$-state automaton $A = \langle Q,\Sigma,\delta \rangle$,
Algorithm~\ref{alg_bidirect} returns the shortest length of a reset word for $A$ or reports that no such a word of length $\leq \mathtt{maxlen}$ exists. 
\end{theorem}
\begin{proof}
In order to prove the correctness of Algorithm~\ref{alg_bidirect}, we introduce additional notation. Let $T^i_c$ denote $T_c$ after performing $i$ steps of BFS, and let $T^j_{ic}$ denote $T_{ic}$ after performing $j$ steps of IBFS. Similarly, let $T^i_v$ denote $T_v$ after performing $i$ steps of BFS, and let $T^j_{vc}$ denote $T_{iv}$ after performing $j$ steps of IBFS. We have the following

\begin{lemma}\label{lm_assignedwords}
For each set $S \in T^i_c$ there is a word $u$ of length $i$, such that $Qu=S$. Similarly for each set $T \in T^j_{ic}$ there is a word $v$ of length $j$, such that $\{q\}v^{-1}=T$ for some state $q \in Q$.
\end{lemma}
\begin{proof} The proof is by induction. For $i=0$ the claim is true with the empty word. For $i>0$,  we note that all new sets $S$ inserted into $T^i_c$ are obtained by applying a letter $a$ to a set $S'\in T^{i-1}_c$ (line~5 of Algorithm~\ref{alg_bfsstep}). By induction hypothesis, there is  a word $u'$ of the length $i-1$ such that $Qu=S'$. Hence, $u'a$ has length $i$ and we have $Qu'a=S'a=S$, as required. The proof of the second statement is dual.
\end{proof}

\bigskip

Let $l$ be the length of the shortest reset words for $A$. First we show that the algorithm in order to report the length of a reset word in line ~19 needs to perform at least $l$ (BFS or IBFS) steps. 

Assume that the algorithm reaches line~19 after $i$ steps of BFS and $j$ steps of IBFS. So there are sets $S \in T^i_c$ and $T \in T^j_{ic}$ such that $S \subseteq T$. By Lemma~\ref{lm_assignedwords}, there are words $u,v$ of lengths $i,j$, respectively, and a state $q\in Q$ such that $Qu=S$ and $\{q\}v^{-1}=T$. Thus, $Quv=\{q\}$, and $uv$ is a reset word of length $i+j$. Consequently, $l \le i+j$.

Now we show that, if $l \le \mathtt{maxlen}$, then the algorithm reaches line~19 after at most $l$ steps. By induction, we prove the following more general statement implying our claim: \emph{for each $i,j \ge 0$, $0 \le i+j \le l$, after $i$ steps of BFS and $j$ steps of IBFS there are sets $S \in T^i_c$ and $T \in T^j_{ic}$, and there exists a reset word $w=uxv$ of length $l$, where $|u|=i,|v|=j,|x|=l-i-j$, such that $Qu = S$ and $\{q\}v^{-1} = T$}.

For $i+j=0$, because of the initialization in lines~5-10, we have that $Q \in T^0_c$ and $\{q\} \in T^0_{ic}$, and a reset word of length $l$\ is as required. Assume that the statement is true for all $i'+j' < i+j$. Assume also, first, that the $(i+j)$-th performed step is BFS one. Then, 
by the induction assumption there exists a reset word $w'=u'x'v$ of length $l$ and sets $S' \in T^{i-1}_c$ and $T \in T^j_{ic}$ such that $Qu' = S'$ and $\{q\}v^{-1} = T$ for some state $q\in Q$, $|u'|=i-1,|v|=j$.

Since $i+j\leq l$, $|x'|>0$. Let $a$ be the first letter of $x'$ and $x'=ax''$. We need to consider two cases, depending on whether $S'a =\delta(S', a)$ (created in line~5 of Algorithm~\ref{alg_bfsstep}) is added (in line~13) into $T^i_c$ or not. If so, then the statement is true, because we have the reset word $w=w'= (ua)x''v$  and sets $S=S' a \in T^i_c$ and $T \in T^j_{ic}$, with required properties..

Otherwise the reason for not adding $S'a$ into $T^i_c$ must be a set $S \in T^i_v$, such that $S \subseteq S'a$ (line~11). Let $u$ be the word corresponding to $S$ by Lemma~\ref{lm_assignedwords}. Then the word $w=ux''v$ (where $x'=ax''$) is a reset word. If $|u| < i$ (we do not know yet if $u\in T^i_c$), then $w$ is shorter than $l$, because $|u|+|x''|+|v| < i + (l-(i-1)-j-1) + j = l$, which is a contradiction. So, $|u| = i$, which means that $S$ has been added into $T^i_v$ in the currently performed $i$-th BFS step. It follows that $S$ has been also added into $T^i_c$. Now, $w=ux''v$ is the required word for $i,j$ with $S \in T^i_c$ and $T \in T^j_{ic}$, $Qu=S$, and $\{q\}v^{-1} = T$.

For the second part of the proof we need to assume that the $(i+j)$-th performed step is IBFS one. In this case the proof is, again, analogous. The difference is that by the induction assumption, we have now a reset word $w'=ux'v'$, and we take into consideration the last letter of $x'$. We leave this part to the reader.
\qed
\end{proof}

\subsection{Finding a Reset Word}\label{subsec_words}

In order to find a reset word of the found minimal length $l$, one needs to apply the following slight modification to the algorithm described above. The main point is that together with the sets stored in the current tries we need to store also the words assigned to these sets. To this end, in line~5 of Algorithm~\ref{alg_bfsstep} (and analogously in the IBFS procedure) we assign to $S'$  the word obtained by concatenating the word assigned earlier to $S$ with the letter $a$ (at the end or at the beginning, respectively). When the goal is reached, the two words are simply merged to form the required reset word. Of course, instead of complete words, with each set we store only a letter and a pointer to the previous part of the word. From these the word is reconstructed when we reach the goal. We note that in this way the asymptotic time and space complexity of the algorithm remain the same.

\section{Heuristics and Optimizations}\label{sec_heuristics} 
In addition to the main part of the algorithm described in the previous section we use a number of heuristics and optimizations. They are justified both by experiments and theoretical arguments. Altogether they can reduce computation time by a factor of at least 25 relative to the implementation without these optimizations. We describe briefly only the most important of them.

\subsection{Estimation of Expected Step Time}\label{subsec_eststep}
To decide which step will be performed in line~12 of the Algorithm~\ref{alg_bidirect} we follow the greedy strategy choosing this step whose execution time, together with the goal test, seems to be smaller at the moment. 
We use a rough estimation of expected execution time by calculating upper bounds for the expected number of visited nodes in subset checking operations, under some simplifying assumptions. Since all other operations in the steps in question are linear in terms of $n$ and the sizes of the current lists, subset checking are the most time consuming operations.
The base for the estimation is the following theoretical result we have established. 
(A set $S \subset X$ is a random subset of $X$ \emph{with Bernoulli distributions} in $[q,r]$ if each element $x$ of $X$ is a member of $S$ with probability $p_x \in [q,r]$.)

\begin{theorem}\label{th_bdist}
Let $p,q,r \in (0,1)$ be such that $q \le r$ and $q > pr$. Let $\mathcal{F}$ be a family of $m$ random subsets of a given set $X$ with Bernoulli distributions in $[q,r]$, and let $S$ be a random subset of $X$ with Bernoulli distributions in $[0,p]$. Then in the trie constructed for the family $\mathcal{F}$, the expected number of visited nodes by the subset checking procedure for $S$ is at most
$$\left(\frac{1+p}{p}+\frac{1}{q-pr}\right) m^{\log_{w}{(1+p)}},$$
where $w = \frac{1+p}{1+pr-q}$.
\end{theorem}
\begin{proof}
Let $f(S,\mathcal{F})$ be the number of visited nodes in the trie constructed for $\mathcal{F}$ by subset checking procedure for $S$. 

Consider the trie constructed for $\mathcal{F}$ as a subtrie of the complete trie. Then $f(S,\mathcal{F})$ can be written as a sum over the nodes in the complete:
$$f(S,\mathcal{F}) = \sum_{x} g(x,S,\mathcal{F}),$$
where $g(x,S,\mathcal{F})$ is an indicator function taking $1$ if the node $x$ is visited and $0$ otherwise. By linearity of expectation,
$$\Ex[f(S,\mathcal{F})] = \sum_{x} \Ex[g(x,S,\mathcal{F})] = \sum_x \Pr(g(x,S,\mathcal{F}) = 1).$$
We can then group the nodes at the same height:
$$\sum_x \Pr(g(x,S,\mathcal{F}) = 1) = \sum_{h = 0}^{\infty} \left( \sum_{\text{$x$ at height $h$}} \Pr(g(x,S,\mathcal{F}) = 1) \right).$$

We will estimate now probability that a node is visited at the height $h$. Let $x$ be a node in the complete trie with the path from the root with exactly $i$ ones and $h-i$ zeros. The node is visited if and only if (1) the searching procedure for a subset of $S$ would reach the node in the complete trie (containing all possible sets) and (2) the node belongs to the constructed trie. These two events are independent, since (1) depends only on $S$ and (2) only on $\mathcal{F}$. We may define therefore two indicator functions: $g'(x,S)$ which takes the value $1$ if the first condition holds (and $0$ otherwise) and $g''(x,\mathcal{F})$ which takes the value $1$ if the second condition holds (and $0$ otherwise).

We bound the probability that condition (1) holds. It holds if and only if $S$ contains all the elements corresponding to ones in the path (otherwise the search does not go into the corresponding branch). Since the probability of containing each element is in $[0,p]$, the probability that the condition (1) holds does not exceed $p^i$. Similarly we bound the probability that condition (2) holds. It holds only if there exists a set in $\mathcal{F}$ whose first $h$ elements correspond to the path of the node (in fact, this condition is necessary, but not sufficient, because of truncating paths). The probability that a single subset has the required sequence of the first $h$ elements, with exactly $i$ ones and $h-i$ zeros, in view of the assumption on Bernoulli distribution in $[q,r]$, can be bounded from above by $r^i(1-q)^{h-i}$. Since $\mathcal{F}$ contains $m$ elements, the probability that condition (2) holds may be upper bounded by $\min\{1,m r^i(1-q)^{h-i}\}$. Summarizing, for a node $x$ with $i$ ones and $h-i$ zeros on the path we have:

\begin{align*}
\Ex[g'(x,S)] & = \Pr[g'(x,S) = 1] = \Pr(\text{$S$ contains the $i$ elements specified by $x$}) \leq p^i, \\
\Ex[g''(x,\mathcal{F})] & = \Pr(g''(x,\mathcal{F}) = 1) = \Pr(\text{$\mathcal{F}$ contains the set encoded by $x$}) \\
& \le \min\{1,mr^i(1-q)^{h-i}\}.
\end{align*}

Now we can group the nodes at the height $h$, which have the same number of ones on the path and we can sum over these groups of the nodes, obtaining:
$$\sum_{\text{$x$ at height $h$}} \Pr[g'(x,S)g''(x,\mathcal{F}) = 1] \le \sum_{i=0}^h \binom{h}{i} p^i \min\{1,m r^i(1-q)^{h-i}\}.$$

This yields a bound that we will use to estimate
$$\Ex[f(S,\mathcal{F})] \le \sum_{h=0}^{\infty} \left( \sum_{i=0}^h \binom{h}{i} p^i \min\{1,m r^i(1-q)^{h-i}\} \right)$$

Let $t = \lfloor \log_{w}m \rfloor$, where $w = \frac{1+p}{1+pr-q}$. We will split up the sum above into two parts: the first one that sums over the levels from $0$ to $t$, and the second one that sums from $t+1$ to $n$.

\textit{Case 1}. We estimate $\sum_{h=0}^{t} \left( \sum_{i=0}^h \binom{h}{i} p^i \min\{1,m r^i(1-q)^{h-i}\} \right)$. For $\Pr[g''(x,\mathcal{F})=1]$ we use in this case the trivial bound $\Pr[g''(x,\mathcal{F})=1] \le 1$. So, we have the bound

$$\sum_{h=0}^{t} \sum_{i=0}^h \binom{h}{i} p^i = \sum_{h=0}^{t} (1+p)^h = \frac{(1+p)^{t+1}-1}{p}.$$

Substituting $t = \lfloor \log_{w}m \rfloor$ yields
\begin{eqnarray*}
\frac{(1+p)^{t+1}-1}{p} & = & \frac{(1+p)^{\lfloor \log_{w}m \rfloor+1}-1}{p} \\
& \le & \frac{(1+p)(1+p)^{\log_{w}(m)}-1}{p} \\
&   = & \frac{(1+p)m^{\log_{w}(1+p)}-1}{p} \\
&   < & \frac{(1+p)}{p}m^{\log_{w}(1+p)}
\end{eqnarray*}

\textit{Case 2}. We estimate $\sum_{h=t+1}^{n} \left( \sum_{i=0}^h \binom{h}{i} p^i \min\{1,m r^i(1-q)^{h-i}\} \right)$. For this case we use the second bound $\Pr[g''(x,\mathcal{F})=1] \le m r^i(1-q)^{h-i}$. We obtain
$$\sum_{i=0}^{h} \binom{h}{i} p^i m r^i (1-q)^{h-i} = m(1+pr-q)^h,$$
 and consequently,
\begin{eqnarray*}
\sum_{h=0}^{\infty} m(1+pr-q)^{h+t+1} & \le & \sum_{h=0}^{\infty} m(1+pr-q)^{h+\log_{w}m} \\
&   = & \sum_{h=0}^{\infty} m\left( m^{\log_{w}(1+pr-q)} (1+pr-q)^h \right) \\
&   = & m^{\log_{w}(w) + \log_{w}(1+pr-q)} \sum_{h=0}^{\infty} (1+pr-q)^h \\
&     & \mbox{(note that $(1+pr-q) < 1$, since by assumption $q>pr$) } \\
&   < & m^{\log_{w}((1+pr-q)w)} \frac{1}{q-pr} \\
&   = & \frac{1}{q-pr} m^{\log_{w}(1+pr-q)\frac{1+p}{1+pr-q}} \\
&   = & \frac{1}{q-pr} m^{\log_{w}(1+p)}
\end{eqnarray*}

Combining both the cases we obtain $$\Ex[f(S,\mathcal{F})] < \frac{1+p}{p}m^{\log_{w}(1+p)} + \frac{m^{\log_{w}(1+p)}}{q-pr} = \left(\frac{1+p}{p}+\frac{1}{q-pr}\right) m^{\log_{w}{(1+p)}},$$
as required.
\qed
\end{proof}

In our empirical observations this optimization reduces computation time by an average of 70\% relative to the implementation performing the BFS and IBFS steps alternatingly. It usually leads to perform slightly more BFS steps, since average sizes of subsets decrease much faster in BFS than increase in IBFS. By a result of Higgins after applying two BFS steps the average size of subsets not greater than $0.55n$ (see \cite{Hi1988}). Our empirical observations show that the two searches meet when the sizes of subsets are as small as $0.03n$. This fact is also the reason why in the goal test we decided to use subset checking of $T_c$ rather than superset checking of $T_{ic}$ (subset checking does not require branching in subtries corresponding to elements not belonging to the queried set).

\subsection{Adding the IDFS Phase}\label{subsec_IDFS}
This is the most important optimization improving not only the performance, but also modifying the general idea.
Bidirectional BFS works if we have no limit on memory resources. Since the number of sets stored in the tries grows exponentially with the number of steps performed, for large automata, we can easily run out of memory. To deal with this, we change the search strategy when we reach the memory limit. Rather than to continue BFS searches we switch to depth-first search, which has restricted memory requirements, and may use the subsets and words computed so far. Moreover, 
assuming the \v{C}ern\'{y} conjecture, we may impose an initial limit on the depth of the search, which allows to make the DFS search \emph{complete}. After each recursive call, when a shorter reset word is found, the limit on the depth of the search is suitably decreased. The search is finished when no limit decreasing is possible and all paths of the limited DFS are exhausted. The search returns either \emph{the shortest reset word} or a counterexample to the \v{C}ern\'{y} conjecture. The IDFS phase is used also to reduce the computation time of the algorithm (even if we are far from reaching the memory limit). This will be discussed in subsection~\ref{subsec_shortcut}.

Our experiments show that it is more efficient to apply the \emph{inverse} DFS, that is, one 
starting from the sets in $T_{ic}$ and computing the preimages to find a set containing a member of $T_c$ (rather than the \emph{forward} DFS starting from the sets in $T_c$ and computing images to find a set contained in a member of $T_{ic}$). An important modification is that we perform search on partial lists of subsets making use of all available memory rather then on single subsets. This gives an additional boost.

\subsection{Reduction of the Automaton}\label{subsec_reduction}
If the input automaton is not strongly connected, after some steps of BFS it can be reduced to a smaller automaton without the states not involved in computation anymore. More precisely, we can remove the states which are not reachable from any state in any subset in the current BFS list. 
So, at the beginning, before the main loop of Algorithm~\ref{alg_bidirect} (line~11), we perform a few steps of BFS and when the size of $T_c$ is larger than $sn$, where $s$ is an experimentally established constant, we check if there are unreachable states in $Q$. This is done by the standard DFS search on $Q$. If this is the case, we create a reduced automaton $A'$ removing the unreachable states, and rebuild all the tries to make them compatible with the reduced automaton. Then, the algorithm may continue using the parameters computed so far. 

Our experiments show that after the first reduction the automaton is usually strongly connected (and no further reduction of this kind can be done). Yet, this optimization is efficient since we have proved that the fraction of strongly connected automata to all automata with $n$ states tends to 0 as $n$ goes to infinity, and that the size of the minimal strongly connected component is on average less than $1-1/e^k$ (provided most automata are synchronizing). From our experiments it follows that for synchronizing automata with $k=2$ this size is $\approx 0.7987n$. Thus, for example, automata with $n=200$ states are reduced on average by as much as 40 states.

\subsection{Reordering of the States}
Efficiency of operations on radix tries depends on the order in which the input automaton's states are processed. We found that the subset checking is performed faster if the states occurring more frequently in queried subsets are later in the ordering. This is because radix tries tends to have logarithmic height (cf. \cite{De1982}), and the states at the end in the ordering are rarely or never checked. As a result, the "effective size" of the queried sets is smaller. To establish frequencies of occurrences of states, and a preferred initial order based on them, we use a stationary distribution of a Markov chain based on the underlying digraph of the automaton. The details will be given in the extended version of the paper.  This optimization is performed before the bidirectional search phase. 

The situation changes completely during the IDFS phase, when the trie $T_c$ is fixed and does not change anymore. The frequencies of occurrences of the subsets in $T_c$ may by computed exactly. This leads to a different reordering. Both reorderings have been confirmed as optimal by experiments. They show that these optimization reduce computation time by an average of 27\%.

\subsection{Using Heuristic Algorithms and IDFS Shortcut}\label{subsec_shortcut}
In order to save a step of search computation we may use known heuristic algorithms to find quickly a good bound for search depth. Therefore, at the beginning of the algorithm, before starting the bidirectional search, we apply a few polynomial time algorithms finding upper bounds for the length of the shortest reset word. In our implementation we use Eppstein algorithm \cite{Ep1990}, FastSynchro algorithm \cite{KRW2012} and our procedure Cut-Off IBFS. The latter is the standard IBFS search with cutting the branches of the search with smallest subsets.
This may spare one step in bidirectional search, if the heuristic algorithms find the shortest word. 

Yet, more importantly, combined with the IDFS phase, this makes possible to reduce the computation time by several orders of magnitude. Knowing that bidirectional search is close to end it is profitable to switch to IDFS phase: at the end the IDFS works much faster, since we do not need to check visited sets and do not need to reconstruct $T_c$ anymore. We call this optimization the \textit{shortcut}. Between steps we use an estimate if it is faster to continue the bidirectional phase or to switch to IDFS phase. Note that the IDFS has a lower constant factor, but the branching factor is equal to $k$. So, it slows the search if started too early. For estimation we use the formula in Theorem~\ref{th_bdist}. Our experiments show that this optimization reduces computation time by as much as 89\%.

\section{Complexity}

The efficiency gain of the algorithm relies mainly on two properties of the majority of automata. First, the average size of subsets decreases fast during the first BFS steps, but increases slow during IBFS steps (cf. subsection \ref{subsec_eststep}). Due to this fact the maintained subsets are usually small. Second, the branching factors of both BFS and IBFS are less than $k$, because of skipping redundant visited sets. Both of the properties are hard to study in a theoretical way, we however have observed them in series of experiments.

To provide a theoretical argument we analyze here the expected running time of the algorithm under some artificial assumptions. We give an upper bound for the bidirectional search only, which is a rough estimate of the expected time, but shows a significant impact of the automata properties on performance. 
The following assumptions are made:
\begin{enumerate}
\item The input is a synchronizing automaton with $n$ states on $k$ letters.
\item The overall branching factor is $r$ in each step of both BFS and IBFS, $1 < r < k$. This corresponds to an effective branching factor, which in view of our experiments is considerably less than $k$.
\item The sets in the tries $T_c,T_v$ and $T_{ic},T_{iv}$ have random Bernoulli distribution:  in each step, they contain any given state with probability $0 < p_c < 1$ (for BFS steps) and $0<p_{ic}<1$ (for IBFS steps). We assume also that $p_{ic} \le p_c$. 
\item The steps of BFS and IBFS are performed alternatingly, starting from BFS.
\item No reductions of the visited tries are made and no IDFS phase is performed.
\end{enumerate}
\indent While the assumptions 2-3 are purely theoretical, they may be treated as an idealization of a typical situation. Using these assumption, denoting by $l$ the length of the shortest reset word of the automaton, we can prove that there exists an integer $0<d<1$, depending on probabilities $p_c,p_{ic}$, such that the following holds.

\begin{theorem}\label{complex}
Under the assumptions {\rm (1-5)} above, and with  $l$ denoting the length of the shortest reset word of the automaton, the expected time complexity of the algorithm is $O(kn^2 r^{l(1+d)/2)})$, and the space complexity is $O(n(k+n)+nr^{l/2})$.
\end{theorem}
\begin{proof}
We use RAM computation model in the analysis, with the uniform cost criteria (that is, each elementary operation costs one time unit). We consider $r,p_c,p_{ic}$ as constants and compute a bound as a function of $n$ and $k$. Let $l$ be the length of the shortest reset word of the automaton. For simplification, assume that $l$ is even.

The initialization phase time may be bounded polynomially by $O(kn^4)$. This includes computing  the inverse automaton $O(nk)$, running the heuristic synchronizing algorithms $O(kn^4)$, computing the stationary distribution $O(n^3)$, changing the order of the states of the automaton $O(nk+n \log n)$, and initializing the tries $O(n^2)$.

Under the assumption on the branching factor, the number of sets in $T_c$ in $i$-th BFS step, after performing $(i-1)$-BFS steps and $(i-1)$-IBFS steps, can be bounded by $r^i$, which is the number of sets after the step. The number of sets in $T_v$ can be bounded by summing added sets during all the BFS steps: $\sum_{j=0}^i r^j = \frac{r^{i+1}-1}{r-1} \in O(r^i)$. Similar bounds hold for $T_{ic}$ and $T_{iv}$, but there are $n$ sets at the beginning, so it yields $O(nr^i)$.

Recall that under assumptions of the theorem we may use Theorem~2, and to obtain the following estimation for the visited number of nodes in the trie
$$\Call{ExpNvn}{m,p,q} = \left(\frac{1+p}{p}+\frac{1}{q-pq}\right) m^{\log_{w}{(1+p)}},$$
where $w = \frac{1+p}{1+pq-q}$. Since we we use this formula for various pairs $p$ and $q$, we shall use an abbreviation $w(p,q) = \frac{1+p}{1+pq-q}$.

Note that each of computing an image or preimage of a set, checking the size of a set,  checking if a set is a subset or superset of another set, can be done in $O(n)$ time. Subset checking for one set can be done in expected time $O(n \Call{ExpNvn}{m,p,q})$, for suitable $m,p,q$. This is so, because we must count not only visited nodes but also test if the set is a subset of a stored set (which costs $O(n)$, but is done at most once for a visited node).

The expected time of the BFS step includes sorting of sets in $L$ (this is done by counting sort, in this case), computing the image of each set by each letter, and checking for visited subsets. So we can bound this by
$$O\left((n r^i) + (k n r^i) + (k n r^i \Call{ExpNvn}{O(r^i),p_c,p_c}) \right).$$
The last component in the sum is dominating, which yields
$$O\left(k n r^i (r^i)^{\log_{w(p_c,p_c)}{(1+p_c)}} \right).$$
Similarly for the bound for the expected time of IBFS step we obtain:
$$O\left(k n r^i (nr^i)^{\log_{w(p_{ic},p_{ic})}{(1+p_{ic})}} \right).$$
Considering the goal test, it is enough to count only the goal test time after the IBFS step (multiplied by $2$). Considering the goal test, in both cases
after the BFS step or IBFS step we can bound the time by
$$
O(n r^i \Call{ExpNvn}{O(r^i),p_{ic},p_c})  =  O(n^2 r^i (r^i)^{\log_{w(p_{ic},p_c)}{(1+p_{ic})}}).
$$
Computing estimated expected step times after $i$-th BFS step and $i$-th IBFS are done in $O(n r^i)$ (having access to list of sets in a trie in linear time), so it may be neglected.

Summing these all yields under domination of the BFS and IBFS step time and the goal test:
\begin{eqnarray*}
& & O\left(k n r^i \left( (r^i)^{\log_{w(p_c,p_c)}{(1+p_c)}} + (nr^i)^{\log_{w(p_{ic},p_{ic})}{(1+p_{ic})}} \right) + n^2 r^i (r^i)^{\log_{w(p_{ic},p_c)}{(1+p_{ic})}} \right) \\
& \in & O\left(k n^2 r^i (r^i)^{d} \right) \\
& =   & O\left(k n^2 (r^i)^{1+d} \right)
\end{eqnarray*}
where $$d = \max((\log_{w(p_c,p_c)}{(1+p_c)}),(\log_{w(p_{ic},p_{ic})}{(1+p_{ic})}),(\log_{w(p_{ic},p_c)}{(1+p_{ic})})).$$
The parameter $d$ depends on the distribution of sets in the tries. Note that $0 < d < 1$, so we could bound $n^d$ simply by $n$.
 
We can now sum over the steps and obtain as the final result the time complexity:
\begin{eqnarray*}
& & \sum_{i=1}^{l/2} O\left(k n^2 (r^i)^{1+d} \right) \\
& \in & O(kn^2 \left(\frac{r^{(1+d)(l/2+1)}-1}{r^{1+d}-1}\right)) \\
& \in & O(kn^2 r^{l(1+d)/2)})
\end{eqnarray*}

The expected space complexity can be bounded by counting stored sets and nodes in the tries after the last step. There are $O(r^{l/2})$ sets in each of the tries. Each set requires $O(n)$ space, also it induces at most $O(n)$ nodes in a trie. The initialization phase can be done in $O(nk+n^2)$ space. So we can state up the space bound for $O(n(k+n)+nr^{l/2})$.
\qed
\end{proof}

We can observe that the expected time is exponential with regard to the length $l$, but the exponent is less than $l$, since $(1+d)/2 < 1$. It is an improvement over the standard BFS algorithm, which has time bound $O(kn R^l)$ (assuming we can check visited sets in constant time). Moreover the standard algorithm usually has a larger branching factor $R > r$, since strict supersets of visited sets are not skipped. The expected space complexity also yields an improvement in comparison to the $O(nR^{l})$ space bound for the standard BFS.

While, generally, our algorithm is exponential in the length $l$ of the shortest reset word, surprisingly, it works fast in polynomial time for the known series of \emph{slowly synchronizing automata}, that is those with $l$ close to the \v{C}ern\'{y} bound. These are automata $\mathrsfs{C}_n$ (the \v{C}ern\'{y} automaton), $\mathrsfs{W}_n$,$\mathrsfs{D}'_n$,$\mathrsfs{D}''_n$, and $\mathrsfs{B}_n$ introduced in \cite{AGV2010}.

\begin{theorem} For the class of the \v{C}ern\'{y} automata $\mathrsfs{C}_n$, and the classes a $\mathrsfs{W}_n$,$\mathrsfs{D}'_n$,$\mathrsfs{D}''_n$, and $\mathrsfs{B}_n$ introduced in {\rm \cite{AGV2010}} the algorithm works in $O(n^4)$ time and $O(n^3)$ space.\end{theorem}

The proof is based on the exact description of the heuristic mentioned in \ref{subsec_eststep}, which shows that for each of the mentioned slowly synchronizing automata the algorithm performs mainly IBFS steps (rather than BFS), and the IBFS lists keep containing only one or two sets (due to reductions of visited subsets).

\section{Experiments}

We performed a series of the following experiments for various $n\leq 320$. For a given $n$, we generate a random automaton $A$ with $n$ states and 2 input letters, check whether $A$ is synchronizing and if so, we find the minimal length of a reset word using the algorithm described in Section~2. On the basis of the obtained results we estimate the expected length of the shortest reset word.

\subsection{Computations}

In the experiments we have used the standard model of random automata, where for each state and each letter all the possible transitions are equiprobable. A~random automaton with $n$ states and 2 input letters can be then represented as a~sequence of $2n$ uniformly random natural numbers from $[0,n-1]$. To generate high quality random sequences we have used the~WELL number generator \cite{PLM2006} (variants 1024 and 19937) seeded by random bytes from \texttt{/dev/random} device. For comparison, recall that Skvortsov and Tipikin, in their experimental study \cite{ST2011}, have generated and checked the following numbers of random automata: 2000 automata for each  $n \in \{1, 2, \ldots , 20, 25, 30, \ldots , 50\}$, 500 automata for each $n \in \{55, 60, 65, 70\}$, and 200 automata for each $n\in \{75, 80, \ldots, 100\}$. In our experiment, up to $7$ states, we have computed exact results checking all automata. For each $8 \le n \le 100$ we checked one million automata, and for each $101 \le n \le 260$ and $n = 265, 270, \ldots, 320$ we checked $10 000$ automata. 
Our computations have been performed mostly on 16 computers with Intel(R) Core(TM) i7-2600 CPU 3.40GHz 4 cores and 16GB of RAM. The algorithm was implemented in C++ and compiled with g++. Distributed computations were managed by a dedicated server and clients applications written in Python.

The average computation time is about $100$ or $1000$ times faster than the time of Trahtman's program TESTAS \cite{Tr2003,Tr2006} for automata with $50$ states. The reduction to SAT used in \cite{ST2011} seemed to be the fastest recently known algorithm and the reported average time for $50$ states automata is $2.7$ seconds, and for $100$ states automata is 70 seconds. Our comparable results are less than $0.006$ and $0.07$ seconds, respectively (we have used faster machines, but only about twice as fast). The Table~\ref{tab:time} presents a rough comparison. The average times are relatively small because of rare occurrences of slowly synchronizing automata. We present also the maximum computation time.

\begin{table}
\centering
\caption{Comparison of average and maximum computation time for random automata.}
\setlength{\tabcolsep}{0.75em}\begin{tabular}{l|cccccc}\hline
$n$                           &  50     &  100     &  150   & 200       & 250         & 300 \\ \hline
TESTAS (\cite{Tr2003})        & 1.4 s   & time-out & --     & --        & --          & -- \\
SAT reduction (\cite{ST2011}) & 2.7 s   & 70 s     & --     & --        & --          & -- \\ \hline
Our average time              & 0.005 s & 0.06 s   & 0.469 s& 2.88 s    & 31.637 s    & 596.249 s   \\
Our maximum time              & 0.26 s  & 3.79 s   & 10.12 s& 159.670 s & 5 h 19 min  & 7 h 55 min \\ \hline
\end{tabular}
\label{tab:time}
\end{table}
\subsection{Results}

Our experiment confirms that for the standard random automata model $\mathbb{A}(n)$ on the binary alphabet the probability that the automaton is synchronizing seems to tend to $1$ as the number $n$ of states grows. 
%%$$P(\textrm{$\mathbb{A}(n)$ is synchronizing}) \xrightarrow[n\to\infty]{}1.$$
This conjecture is posed in \cite{ST2011}, but we have heard it earlier from Peter Cameron during BCC conference in Exeter 2011. For $n=100$, $2250$ of one million automata turned out to be non-synchronizing ($0.225 \%$), and for $n=300$, only five of $10000$ automata. The graphical representations of our experiments in this respect forms a smooth curve very fast converging to $1$. We observe also that random automata mostly are not strongly connected. 

The main result of our experiments is the estimation of the expected length of the shortest reset word. We deal with the infinite sequence of random variables $\ell(n)$ defined as the length of the shortest reset word for a random synchronizing automaton with $n$ states. 
We have observed that the approximation $\E[\ell(n)] \approx 1.95n^{0.55}$ proposed in \cite{ST2011} is inflated. Based on currently available data, we propose a new more precise experimental approximation for the expected length $\E[\ell(n)] \approx 2.5\sqrt{n - 5}.$
A comparison of the estimations with the experimentally obtained mean length is given in Figure~1.
We observe also that our result suggest that the expected length may belong to $\Theta(\sqrt{n})$.

In contrast with the experiments by Skvortsov and Tipikin \cite{ST2011}, our experiments allow also to obtain a good estimation of the approximation error. Making use of the well-known Hoeffding's inequality, we obtain the following:

\begin{theorem} 
Let $ML(n)$ denotes the mean length of the shortest reset word of the automata in the sample of $m$ randomly generated  synchronizing $n$-state automata. If the ratio of the automata with the length of the shortest reset word larger than $M_n$ to all automata in the sample does not exceed $r$, then with probability at least $1-p$ 
$$|ML(n) - \E[\ell(n)]| \leq M_n(1-r)\sqrt{\frac{\log (2/p)}{2m}} + \frac{n^3}{6}r.$$
\end{theorem} 
\begin{proof}
We make use of the well-known Hoeffding's inequality \cite{Hoeffding1963}. Given $0 < p \leq 1$, with probability at least $1-p$
\begin{equation}
|\overline{X}-\Ex[\overline{X}]| \leq  R \sqrt{\frac{\log (2/p)}{2m}},
\end{equation}
where $\overline{X} = (X_1+\ldots+X_m)/m$ is the empirical mean of random variables $X_1,\dots,X_m$ with the same range $R$. 
Since the distribution of $\ell(n)$ is highly asymmetric, one needs to combine this inequality with the statistical fact that the maximal lengths of the shortest reset words obtained in the experiment are much smaller than the known bounds and that longer lengths occur rarely.

Let $M_n$ be the maximal length of the shortest reset word for the $n$-state automata generated in the experiment and $m$ the size of the sample. 
First we assume that we sample only automata with the length $\leq M_n$. Denote the corresponding random variable by $\ell'(n)$.
Applying the Hoeffding's inequality, putting $X_1=\ldots =X_m = \ell'(n)$, and $R=M_n$  we obtain   
$$
|ML(n)-\Ex[\ell']| \leq  M_n \sqrt{\frac{\log (2/p)}{2m}}.
$$
Let $\ell''(n)$ be the length of the shortest reset word for a synchronizing automata with $n$ states $ \ell(n) \geq M_n$. Then  we obtain
$$
|ML(n)-\Ex[\ell]| \leq (1-r)|ML(n)-\Ex[\ell']| + r|ML(n)-\Ex[\ell'']|\leq  (1-r)M_n \sqrt{\frac{\log (2/p)}{2m}}+ r\frac{n^3}{6},
$$
as required.
We have used the well-known bound $n^3/6$ for the length of the shortest reset word.
\qed
\end{proof}

\begin{figure}[t]
 \centering
 \epsfig{file=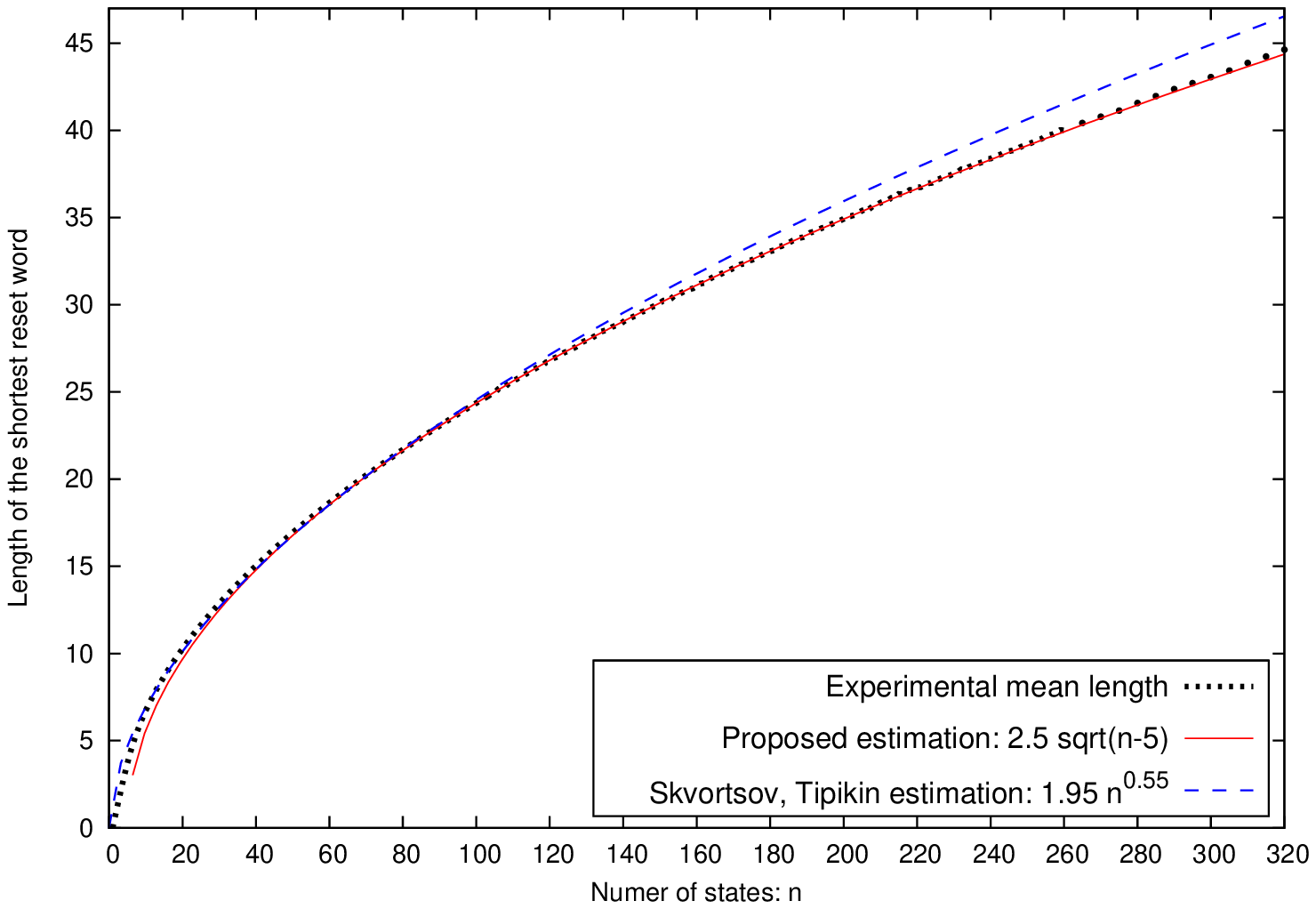, width=100mm, angle=-90}
 \caption{Experimental mean length of the shortest reset words compared with estimations.}
\end{figure}
Assuming the \v{C}ern\'{y} conjecture in the last term $n^3/6$ may be replaced by $(n-1)^2$ (giving essentially better estimation). Let us take $n=100$, $m=10^6$ and $p=0.0001$. 
Since, with probability $q=(1-r)^m$ the ratio of the automata with the shortest reset word longer than $M_n$ is less than $r$, one may see that for $1/r\geq 100975$, $q < 0.0001$. Hence, with high probability $1/r> 100975$, and taking into account the experimental value $M_{100}=41$, 
the error is less than 1.75 (or 0.19 assuming  the \v{C}ern\'{y} conjecture). This means that with high probability the expected length of the shortest reset word for synchronizing automata with $n=100$ states is close to our experimental result $ML(100)=24.34$. Comparing this with the results of Skvortsov and Tipikin \cite{ST2011}, we note that, for automata with $100$ states, they also have obtained the expected length close to $24$,  but the small size of their sample $m=200$ does not allow any reasonable estimation of the error. Other interesting claims of \cite{ST2011} concerning the variance and approximation of $\ell(n)$ will be discussed in the extended version of the paper.

%%%%%%%%%%%%%%%%%%%%%%%%%%%%%%%%%%%%%%%%%%%%%%%%%%%%%%%%%%%%%%%%%%%%%%%%%%%%%%%%%

%%%%%%%%%%%%%%%%%%%%%%%%%%%%%%%%%%%%%%%%%%%%%%%%%%%%%%%%%%%%%%%%%%%%%%%%%%%%%%%%%
\end{document}